\documentclass[
aps,%
12pt,%
final,%
notitlepage,%
oneside,%
onecolumn,%
nobibnotes,%
nofootinbib,%
superscriptaddress,%
noshowpacs,%
centertags]%
{revtex4}

\begin {document}
\begin{center}
\large \bfseries \MakeTextUppercase{A Spectroscopic Study of the Envelope of the Hybrid Nova V458 Vul and the Surrounding Planetary Nebula}
\end{center}
\author{\firstname{T. N.}~\surname{Tarasova}}

\email{taya_tarasova@mail.ru}
\affiliation{Crimean Astrophysical Observatory, Nauchnyi, Crimea, Russia}

\begin{abstract}
Spectroscopic observations of the hybrid V458 Vul obtained between days 9 and 778 after the brightness maximum are analyzed. Short-period, daily profile variations of forbidden [FeVII] iron lines were detected in the nebular phase, as well as a long-period (about 60-day) cyclic variation that was correlated with the photometric and X-ray cycles. The abundances of helium, neon, and iron in the nova's envelope have been estimated. The helium, neon, and iron abundances exceed the solar values by factors of 4.4, 4.8, and 3.7. The envelope mass is 1.4$\times$ 10$^{-5}$~M$_{\odot}$. The electron temperatures and number densities have been calculated for the Northwestern and Southeastern knots of the planetary nebula. The temperature derived for the Northwestern knot is Te = 10 000 K and the electron number density, n$_{e}$ = 600 cm $^{-3}$ for the Southeastern knot, Te = 13 000 K and n$_{e}$ = 750 cm$^{-3}$.
\end{abstract}

\maketitle
\section{Introduction}
This paper presents a study of the envelope of V458 Vul, which is a member of the rare class of hybrid novae. Only a few stars of this kind are currently known. Apart from V458~Vul, the class includes LMC 1988-2, V3890 Sgr \cite{Williams1991}, V838 Her \cite{Williams1994}, M31 N 2006-10b \cite{Shafter2011}, and the recurrent nova T Pyx \cite{Shore2011}. V458 Vul was discovered as a nova on August 8, 2007, when its magnitude was 9.5$^{m}$ \cite{Nakano2007}. Within a day, it had reached its maximum, m$_{V}$ = 8.1$^{m}$. The name V458 Vul was given to the nova by Samus \cite{Samus2007}. An interesting feature of this nova is that it lies inside a planetary nebula \cite{Wesson2008}. Only one such nova, GK Per 1901, was known previously \cite{Bode1987}. The existence of such novae confirms that the central stars of planetary nebulae can be binary systems. 

Buil and Fujii \cite{Buil2007} and Munari \cite{Munari2007} obtained the first low-resolution spectra of V458~Vul one day after its discovery. In addition to Balmer HI lines, the spectra contained lines of ionized iron, FeII. All the spectral lines had P Cygni profiles. For this reason, the star was classified as a FeII nova, according to the spectral classification introduced by Williams \cite{Williams1991,Williams1992}. However, subsequent observations by Skoda \cite{Skoda2007} and Tarasova \cite{Tarasova2007} showed a strengthening of the HeI lines and the absence of lines with P Cygni profiles in the spectrum; i.e., the nova more closely resembled a HeN nova at that stage. We classified it as a hybrid nova \cite{Tarasova2007}. This classification was confirmed by Poggiani \cite{Poggiani2008}. Infrared spectroscopy carried out on October 10--11, 2007 \cite{Lynch2007} also indicated abnormally strong neutral-helium lines, confirming that the type of nova V458 Vul had changed.

We present here spectroscopy covering the interval between days 9 and 778 after the nova outburst. Some of these observations were obtained during the nova's brightening by almost 1.5$^{m}$. Brightenings were observed during the smooth fading phase, as well as the nebular phase. We detected variations of the line profiles in spectra taken during the brightenings, and analyzed this variability. Spectra taken during the late nebular phase indicated line radiation from the symmetrically located knots of the planetary nebula. We used these lines to derive the physical characteristics of the corresponding structures. Spectra taken during the nova's nebular phase have a very strong HeII 4868 line, along with forbidden lines of iron, [FeVII], and neon, [NeV]. We estimated the abundances of these elements in the nova envelope and the mass of the envelope.

\section{LIGHT CURVE OF V458 Vul}

Figure 1 presents a light curve of V458 Vul constructed using the AAVSO database. The insert in this figure shows the star's fading during the first 20 days. The smooth fading was superposed with
temporary brightenings by 1.5$^{m}$ on days 4 and 10 after the maximum. The duration of the first brightening was about three days, and that of the second one about four days, with the interval between them being about three days. Moreover, one year after the brightness maximum, the nova displayed variability both within a day and over longer periods. The last longlived (about 60 day) brightening occurred during the phase of the final brightness decline. The star's brightness first increased by almost 2$^{m}$ over approximately 30 days, then decreased by the same amount over the same time between days 330 and 390. A similar brightening occurred between days 440 and 500. Apparently, the system exhibited cyclic brightness variations, with a cycle duration of about 60 days.

\begin{figure}
\hspace{-35mm}
\includegraphics[scale=0.6]{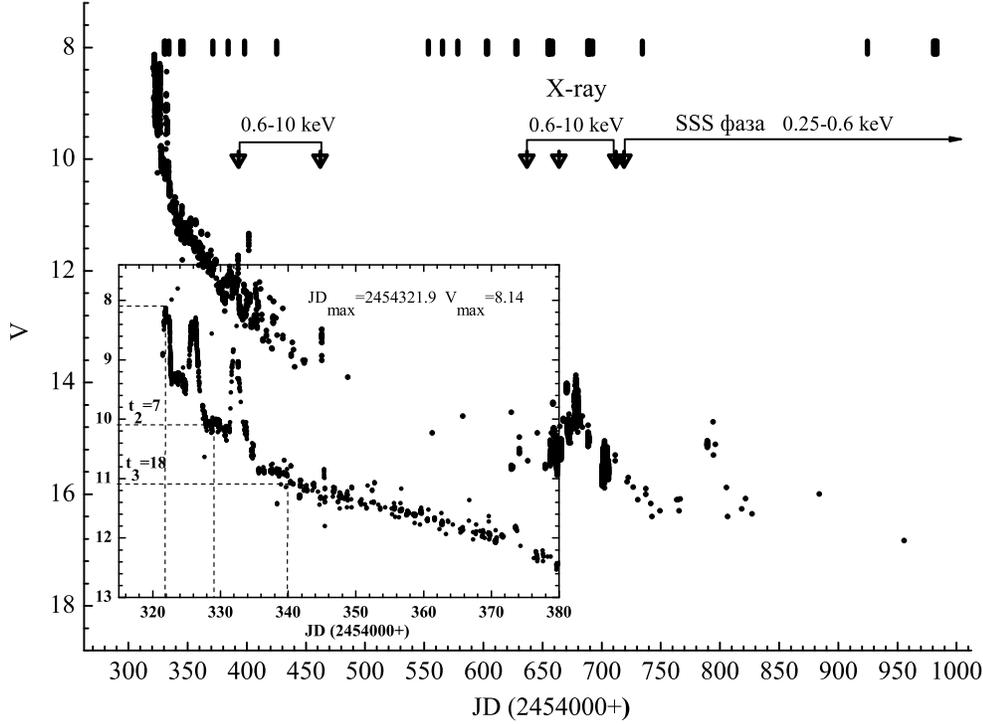}
\caption{Light curve of V458 Vul for 2007--2009. The observations were retrieved from the AAVSO database}
\end{figure}

We used the smoothed light curve to determine the parameters t$_{2}$ and t$_{3}$, which were found to be 7 and 18 days, respectively \cite{Tarasova2008}. These parameters give the times for the star to fade by 2$^{m}$ and 3$^{m}$. According to these values, the nova is a fast novae according to the classification of Payne-Gaposchkin \cite{Payne-Gaposchkin1957}. Using statistical relations between the absolute magnitude and the times t$_{2}$ and t$_{3}$ \cite{Downes2007}-\cite{Cohen1985}, we calculated the absolute magnitudes at the maximum brightness -9.2, -8.8, and -8.7. Using these absolute magnitude, we estimated the distance to the nova to be from 10.1 to 12.7 kpc, depending on the absolute magnitude value. 

Our estimate of the white dwarf's mass from the relation M$_{max}$(V)=-8.3-10log(M$_{wd}$/M$_\odot$) \cite{Livio1992}, M$_{wd}$ = 1.3 M$_\odot$, indicates that V458 Vul may be a candidate Type Ia supernova.

\section{SPECTRAL EVOLUTION AND SPECTRAL LINE PROFILE VARIATIONS OF V458 Vul}

Our spectroscopic observations were performed with the SPEM spectrograph at the Nasmyth focus of the 2.6-m Shajn reflecting telescope of the Crimean Astrophysical Observatory. With two exceptions, all the spectra were taken with a dispersion close to 2 \AA\ per pixel. For two of the spectra, the mean dispersion was 0.75 \AA\ per pixel. We reduced the spectra in the standard way, taking into account bias, flat-fielding and calibration of the wavelength scale. We tied the spectra to the wavelength scale using the spectrum of a helium-neon lamp. We calibrated the fluxes in the star's spectrum based on the absolute spectral energy distribution of the spectrophotometric standard HR 7679, taken from the catalog of Burnashev \cite{Burnashev1985}. We observed the spectrophotometric standard on the same date as V458 Vul and at the same zenith distance, so that we had no need to introduce corrections for the air-mass difference between the standard star and the nova. Since we used a slit spectrograph, for check purposes, we compared the B and V color indices to those calculated from the calibrated spectra taken on the same dates as the photometric magnitudes. On average, the difference between the calculated and measured magnitudes was about 0.1$^{m}$.

The light curve in Fig. 1 has notes for the dates of the spectroscopic observations; Table~1 presents a log of the observations, with information on the nova's spectroscopic data. We present a detailed description of the nova's spectral evolution below, based on subdividing the entire time interval of our observations into three phases and relatng the phases to particular parts of the nova's light curve: the phase of initial fading (days 9--25 after the brightness maximum), the transition phase (days 49--104 after the brightness maximum), and the longest phase, which is the phase of the final brightness decline, which is accompanied by the appearance of forbidden lines in the nova's spectrum. This final phase, days 232--778 after the maximum, covers an interval of about 1.5 years.

\begin{table}[t]
\centering{\bf Table 1. Log of observations of V458 Vul}
\begin{tabular}{|c|c|c|c|c|c|c|c|}
\hline\hline
Date &	JD    & Days  &  Spectral      &Date &	JD        & Days &  Spectral \\
     &2450000+&       &   range, \AA   &     &  2450000+  &      &     range,   \AA \\
\hline
17.08.2007 &4330.255  &  9           &4225-5175      &11.07.2008 &4658.246  &  337         &3950-7574\\
18.08.2007 &4331.282  &  10          &3625-7575      &11.07.2008 &4658.406  &  337         &3950-7574\\
21.08.2007 &4334.257  &  13          &3725-7624      &10.08.2008 &4688.369  &  367         &3774-7624\\
31.08.2007 &4344.271  &  23          &3784-7624      &10.08.2008 &4688.542  &  367         &3375-7125\\
02.09.2007 &4346.310  &  25          &3775-7623      &11.08.2008 &4689.320  &  368         &3374-7225\\
27.09.2007 &4371.232  &  49          &3776-7524      &11.08.2008 &4689.488  &  368         &3374-7149\\
10.10.2007 &4384.298  &  63          &3800-7575      &12.08.2008 &4690.317  &  369         &4841-7075\\
23.10.2007 &4397.185  &  76          &3996-7525      &12.08.2008 &4690.364  &  369         &3349-7100\\
20.11.2007 &4425.168  &  104         &3825-7575      &12.08.2008 & 4690.363 &  369         &3749-6175\\
28.03.2008 &4553.582  &  232         &3724-7574      &14.08.2008 & 4692.367 &  371         &3349-7575\\
09.04.2008 &4565.524  &  243         &4950-7300      &14.08.2008 &4692.508  &  371         &3349-7099\\
21.04.2008 &4578.468  &  257         &3275-5475      &16.08.2008 &4694.333  &  373         &4075-5050\\
21.04.2008 &4578.468  &  257         &5925-7575      &25.09.2008 &4734.310  &  413         &3349-7574\\
15.05.2008 &4602.502  &  281         &3350-7574      &03.04.2009 &4924.563  &  603         &4549-6925\\
16.05.2008 &4603.506  &  282         &3750-7574      &29.05.2009 &4980.459  &  659         &4700-7624\\
10.06.2008 &4627.543  &  306         &3800-6815      &30.05.2009 &4981.461  &  660         &3770-5687\\
11.06.2008 &4628.437  &  307         &3800-7575      &31.05.2009 &4982.482  &  661         &4700-6839\\
06.07.2008 &4654.437  &  333         &3800-7624      &01.06.2009 &4983.493  &  662         &3324-6850\\
08.07.2008 &4655.372  &  334         &3776-7575      &29.06.2009 &5011.446  &  690         &5200-7450\\
08.07.2008 &4655.538  &  334         &3374-7575      &01.07.2009 &5013.468  &  692         &4750-7024\\
09.07.2008 &4654.344  &  335         &3775-7574      &26.08.2009 &5070.412  &  749         &4724-6850\\
09.07.2008 &4656.521  &  335         &3800-7575      &24.09.2009 &5099.384  &  778         &3594-6850\\
\hline 
\end{tabular}
\end{table}

\subsection{Early Spectral Evolution of the Nova (Days 9--25)}

The spectra corresponding to this phase, taken on days 9, 10, 13, 23, and 25 after the maximum, are Fig. 2. The first and second spectra are dominated by lines of HI and FeII (the 27, 28, 37, 38, 42, 49, and 74 multiplets). The presence of these lines is characteristic of novae of the FeII spectral type. However, along with hydrogen lines, helium lines become the strongest in the later spectra, indicating an HeN spectral type. We accordingly classified this as a ``hybrid'' nova. The envelope expansion velocity measured on day 10 from the H$\alpha$ and H$\beta$ lines was 1450 km/s, while the velocity measured from the HeI lines was 1560 km/s.

\begin{figure}
\hspace{-20mm}
\includegraphics[scale=0.8]{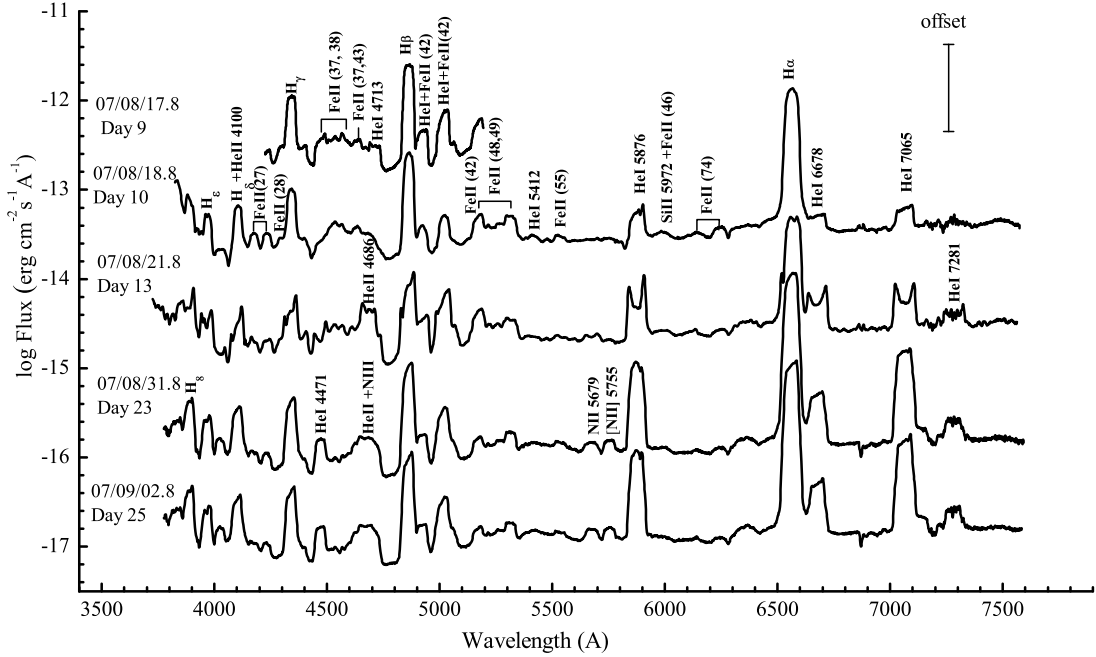}
\caption{Low-resolution spectra taken during the phase of rapid brightness decline during days 9-25 after the brightness maximum. The third spectrum was obtained just after the second brightening. For clarify of presentation, each spectrum has been moved relative to the previous one by equal shifts.}
\end{figure}

The third spectrum in Fig. 2 (for day 13), taken after the second brightening, is very different from the others: the profile shapes of the neutral hydrogen and helium lines have changed cardinally. The neutral hydrogen line profiles now have two components, and the helium line profiles are two-peaked or saddle-shaped. The blue and red peaks of the three-component HI line profiles have radial velocities of -2000$\pm$50 and 1400$\pm$50 km/s, respectively. The radial velocity of the central component is about 200--250 km/s. The radial velocities of the blue and red peaks of the two-peaked HeI line profiles are -1800$\pm$50 and 1600$\pm$50 km/s. The three-component HI profiles are probably a superposition of two profiles -- one rounded and one saddle-shaped. Support for this hypothesis is that the distance between peaks to the blue and the red side of the HI and HeI line profiles is the same, about 3400 km/s. Figures 3 and 4 display the HI and HeI profiles. Similar profiles were found by Mauclaire \cite{Mauclaire2007}, Skoda et al. \cite{Skoda2007}, and Rajabi et al. \cite{Rajabi2012} after the first brightening. Ragan et al. \cite{Ragan2010} also observed saddle-shaped profiles of the HeI 6678 line. Based on the fact that the profile shapes were virtually the same after the first and second brightenings, we suggest that similar changes in the envelope structure were observed at these phases.

Saddle-shaped profiles are usually observed if a system contains an accretion disk or undergoes circumpolar ejections of matter from the white dwarf's surface. Rajabi et al. \cite{Rajabi2012} considered a model with an ejection in the form of a disk, using interferometric observations at 2.2 $\mu$m for their analysis.
\begin{figure}
\hspace{-35mm}
\includegraphics[scale=0.6]{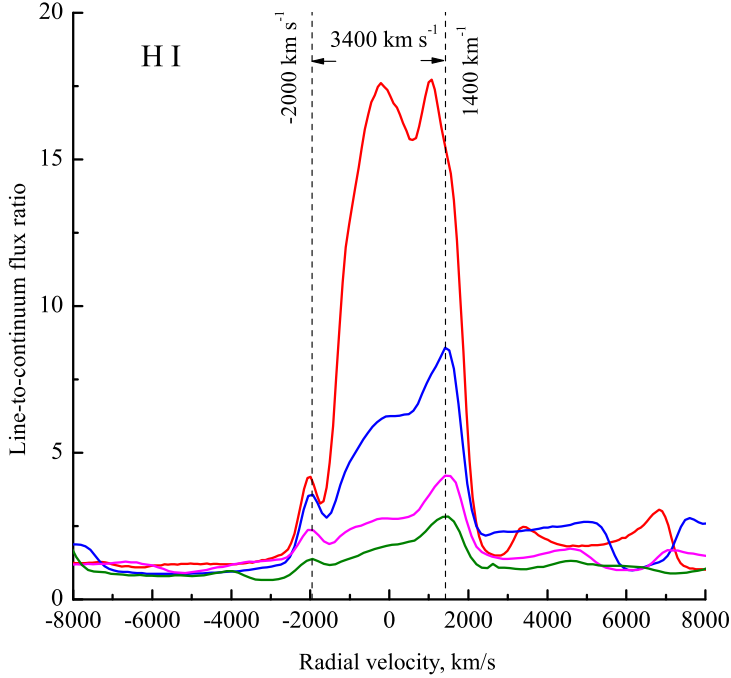}
\caption{Profiles of the Balmer H$\alpha$, H$\beta$, H$\gamma$, and H$\delta$ lines. The fluxes in the lines were normalized to those in the continuum.}
\end{figure}
\begin{figure}
\hspace{-35mm}
\includegraphics[scale=0.6]{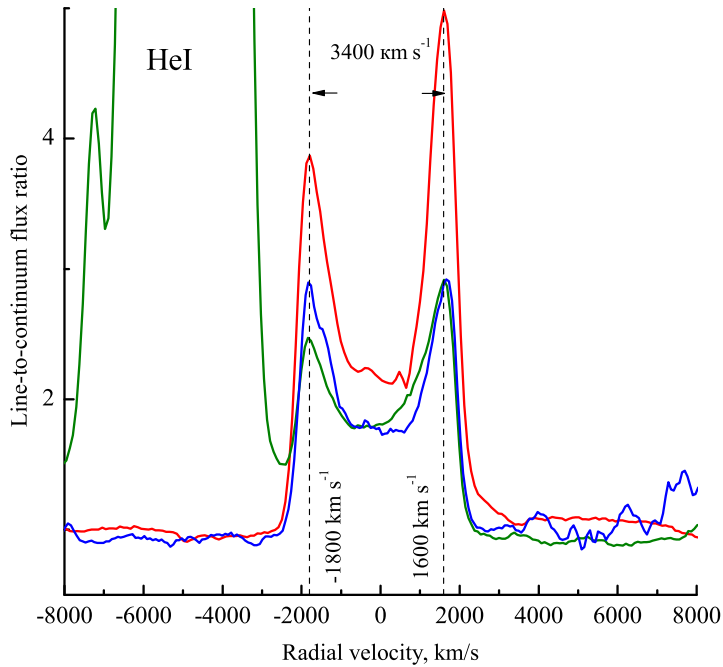}
\caption{Profiles of the HeI 5876, 6678, and 7065 spectral lines. The fluxes in the lines were normalized to those in the continuum.}
\end{figure}

Our spectra obtained after the second brightening suggest that the second ejection was identical to the first: the line profiles observed after the second brightening were exactly the same as those after the first brightening. The hypothesis that this is associated with matter ejections is supported by the P Cygni line profiles in the spectra after both the first \cite{Rajabi2012} and second \cite{Mauclaire2007} brightenings. We suggest that these were short-term matter ejections, additional to the main ejection of the envelope, since the line profiles changed several days later.

On days 24 and 25 after the brightness maximum, the HeI and HI profiles again changed: the HeI lines returned to their pre-brightening shapes, and the HI lines began to resemble the HeI lines. The envelope expansion velocity had decreased by days 24 and 25 to 1370~km/s for the  H$\alpha$ and H$\beta$ lines and 1390 km/s for the HeI lines.

\subsection{Spectral Variations in the Transition Phase (Days 49--104)}

The spectra corresponding to the transition phase are presented in Fig. 5. In this phase, the FeII lines disappear, and the HeI lines become stronger, to the extent that the intensities of the strongest (HeI 5876 and 7065) are virtually the same as that of H$\beta$. The blend consisting of FeII ionized-iron lines is replaced with the NII + NIII + HeII 4686 blend in the 4570-4740 \AA\ region. Moreover, the NII 5679, [NII] 5755, and [ArIII] 7136 lines appeared in the spectrum. The derived envelope expansion velocities decreased to 1360 km/s (HI lines) and 1390 km/s
(HeI lines). Absorption components appeared in the line profiles by days 49 and 63 after the brightness maximum. The radial velocity of these components derived from the HeI 5876 line was about 3100 km/s. Apparently, thermonuclear reactions continuing on the white dwarf gave rise to chaotic gas ejections, with absorption components indicating the presence of matter outflowing from the white dwarf appearing in the spectral lines. Approximately at this time, between days 70 and 140, X-ray emission was detected \cite{Tsujimoto2009}. The line profiles also changed in this stage, but these variations were not as substantial as during the brightenings.

\begin{figure}
\hspace{-20mm}
\includegraphics[scale=0.8]{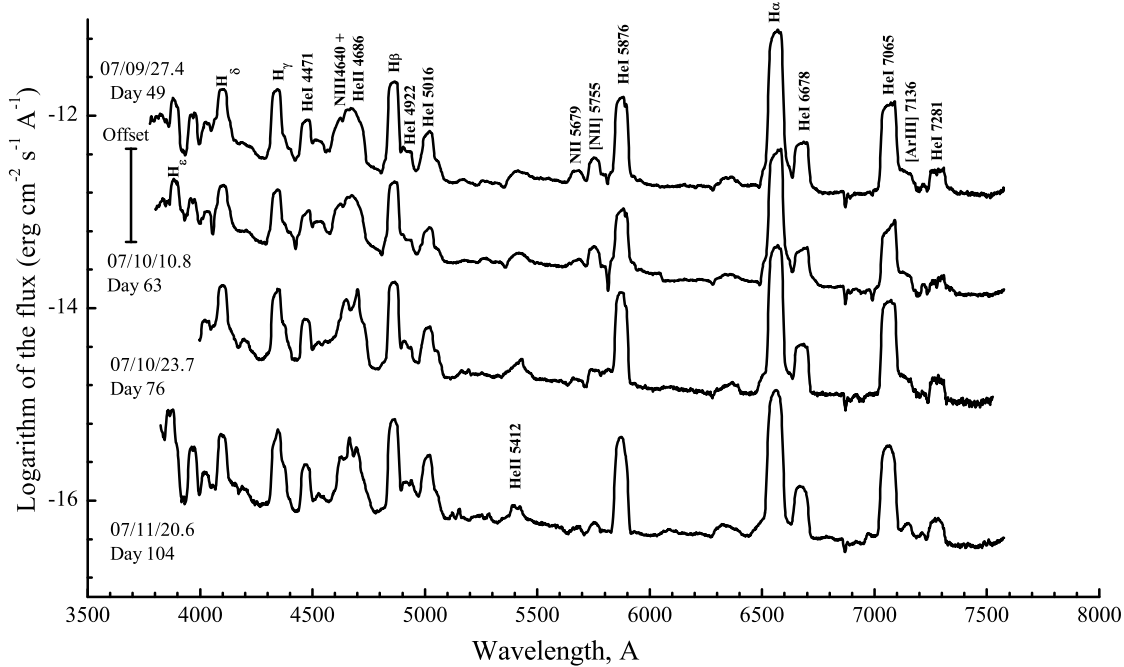}
\caption{Low-resolution spectra obtained during the transition phase, during days 49-104 after the brightness maximum. The presentation of the spectra is the same as in Fig. 2.}
\end{figure}

\subsection{Spectroscopic Observations of V458 Vul in the Forbidden-Line Phase (Days 232--778)}

Figure 6 shows that the forbidden-line phase developed between days 104 and 232. The mean envelope expansion velocity derived from the H$\alpha$ and H$\beta$ lines decreased to 1300 km/s. Apart from HI lines, the strongest lines in the spectrum are now HeII 4686, [FeVII] 6086, and [NeV] 3426. The spectrum also contained lower-intensity [FeVII] 5721, [FeVI] 5176, and [CaV] 5309 lines, as well as the [FeVII] 4942 + [FeVI] 4974 blend. The spectrum also contained weaker HeII 5412, [FeVII] 3586, 3759, [FeVII] and [ArV] 7006 lines. Starting from day 413, the [FeX] 6375 line appeared in the spectrum, and narrow lines belonging to the planetary nebula became appreciable. The presence of forbidden lines of iron in high ionization states [FeVI] and [FeVII], and of neon [NeV], confirms our classification of V458 Vul as a hybrid nova. A distinguishing characteristic of the nova is an absence of [OIII] 4363, 4959, 5007 lines, which become the strongest lines for most novae at this stage. In the case of V458 Vul, these lines are either completely absent or very weak, and are blended with the HeI 4922, 5016, [FeVI] 4967, 4972, and [FeVII] 4989 lines. A similar spectrum was also observed for the nova V2214 Oph \cite{Williams1994}.

\begin{figure}
\hspace{-20mm}
\includegraphics[scale=0.8]{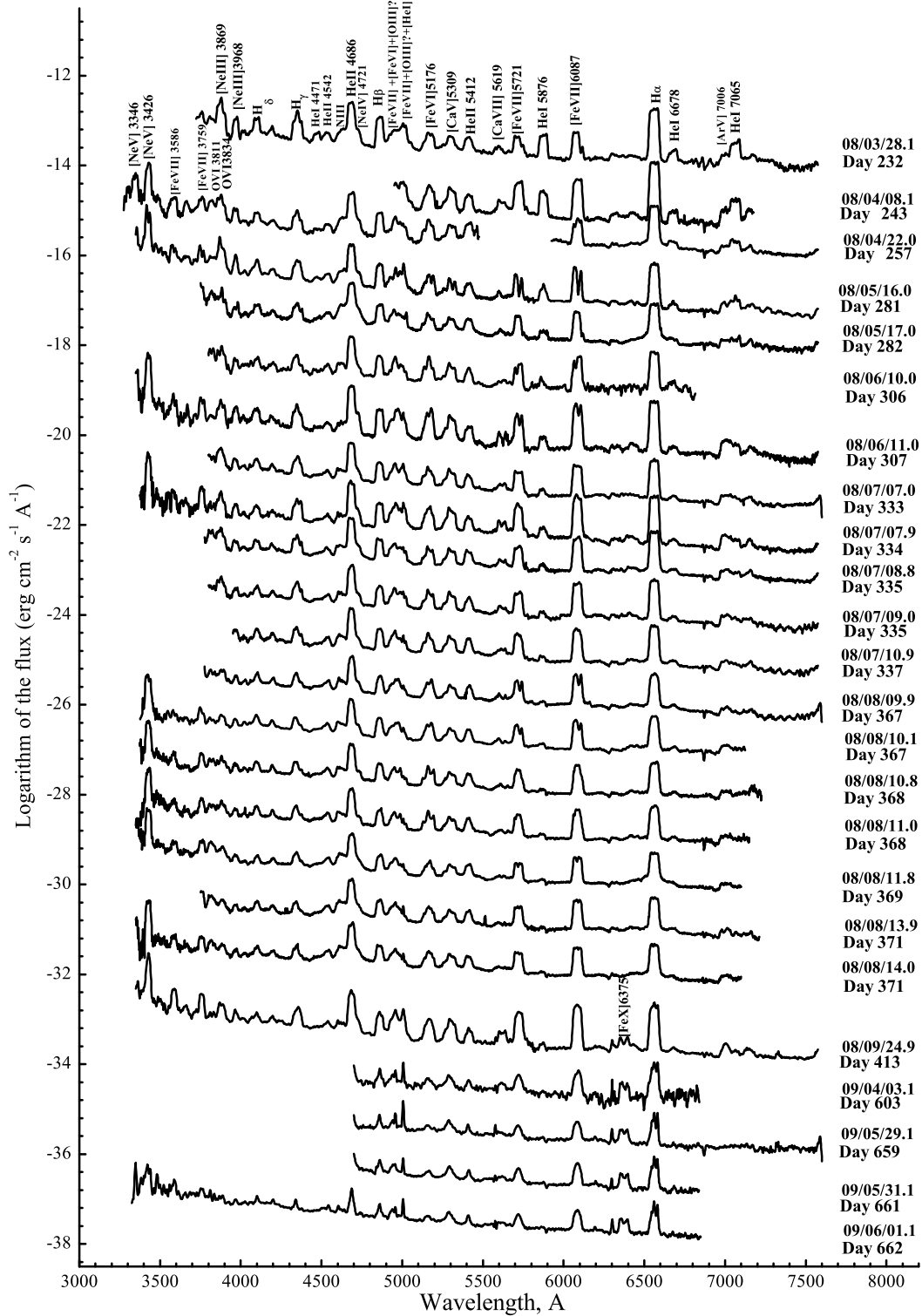}
\caption{Low-resolution spectra obtained during the forbidden-line phase (days 232--662 after the brightness maximum). The presentation of the spectra is the same as in Fig. 2.}
\end{figure}

In this phase, on day 330, the nova became brighter by more than 1$^{m}$. The change in brightness lasted for almost 60 days (approximately from day 330 to day 390). The brightening maximum occurred approximately on day 360. The light curve corresponding to this phase is displayed in Fig. 7. The times when spectra were obtained are marked on the light curve. The time interval when the X-ray flux from the star was measured is also indicated. The star had not yet entered the phase of soft X-ray emission at this time. A harder, variable X-ray flux that was correlated with the optical brightness was detected at 0.6-10 keV \cite{Tsujimoto2009}. We found strong variations of the [FeVII] forbidden-line profiles in this stage. Other lines also varied, but the strongest profile variations were detected for the [FeVII] lines (see Fig. 9 below).

\begin{figure}
\hspace{-20mm}
\includegraphics[scale=0.8]{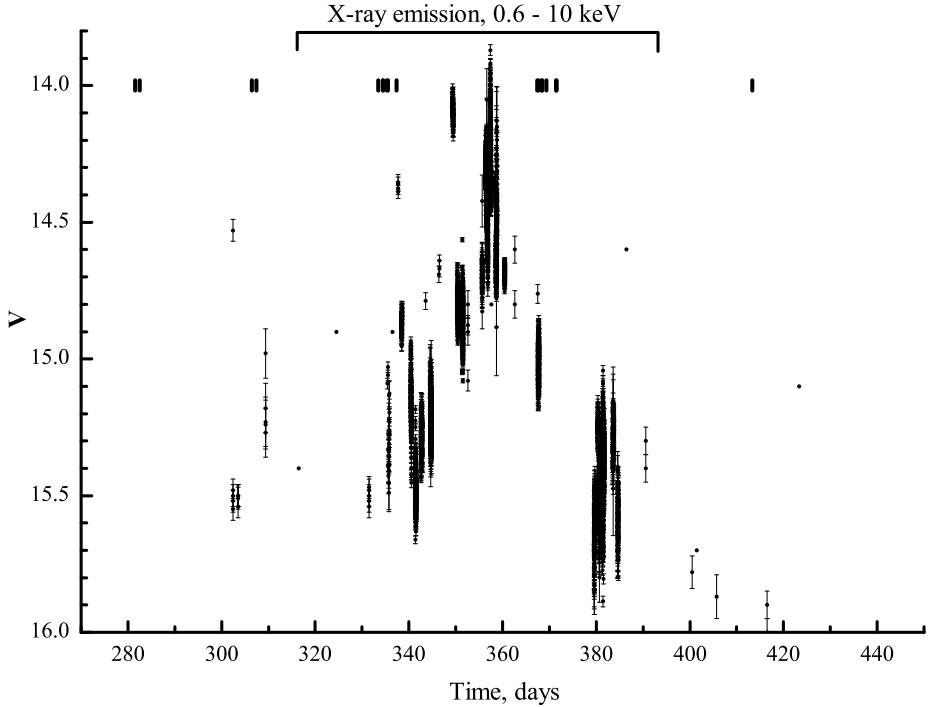}
\caption{Fragment of the light curve during the forbidden-line phase (days 232--662 after the brightness maximum).}
\end{figure}

\subsection{Variability of the [FeVII] Profiles.}
For comparison, Figs. 8 and 9 display profiles of the H$\alpha$, HeII 4686, and [FeVII] 6087 lines. The profiles of the H$\alpha$ and HeII 4686 lines differ from the [FeVII] 6087 forbidden-line profile, with the exception of the profiles recorded on day 413. The [FeVII] 6087 profiles display daily variations (from day 281 to day 282, Fig. 8; from day 333 to day 337 and from day 367 to day 371, Fig. 9). Note, however, that some changes in the [FeVII] 6087 line profile occurred within four hours (days 367, 368; Fig. 9). The [FeVII] 6087 profiles recorded on days 281, 306, 307, and 367 show deep dips, with the positions and depths of the dips on days 307 and 367 being nearly identical (Fig. 9). The time interval between the repeated profiles (60 days) is the same as the duration of the cycle of the photometric and X-ray variations \cite{Ness2009}.

\begin{figure}
\hspace{-20mm}
\includegraphics[scale=0.8]{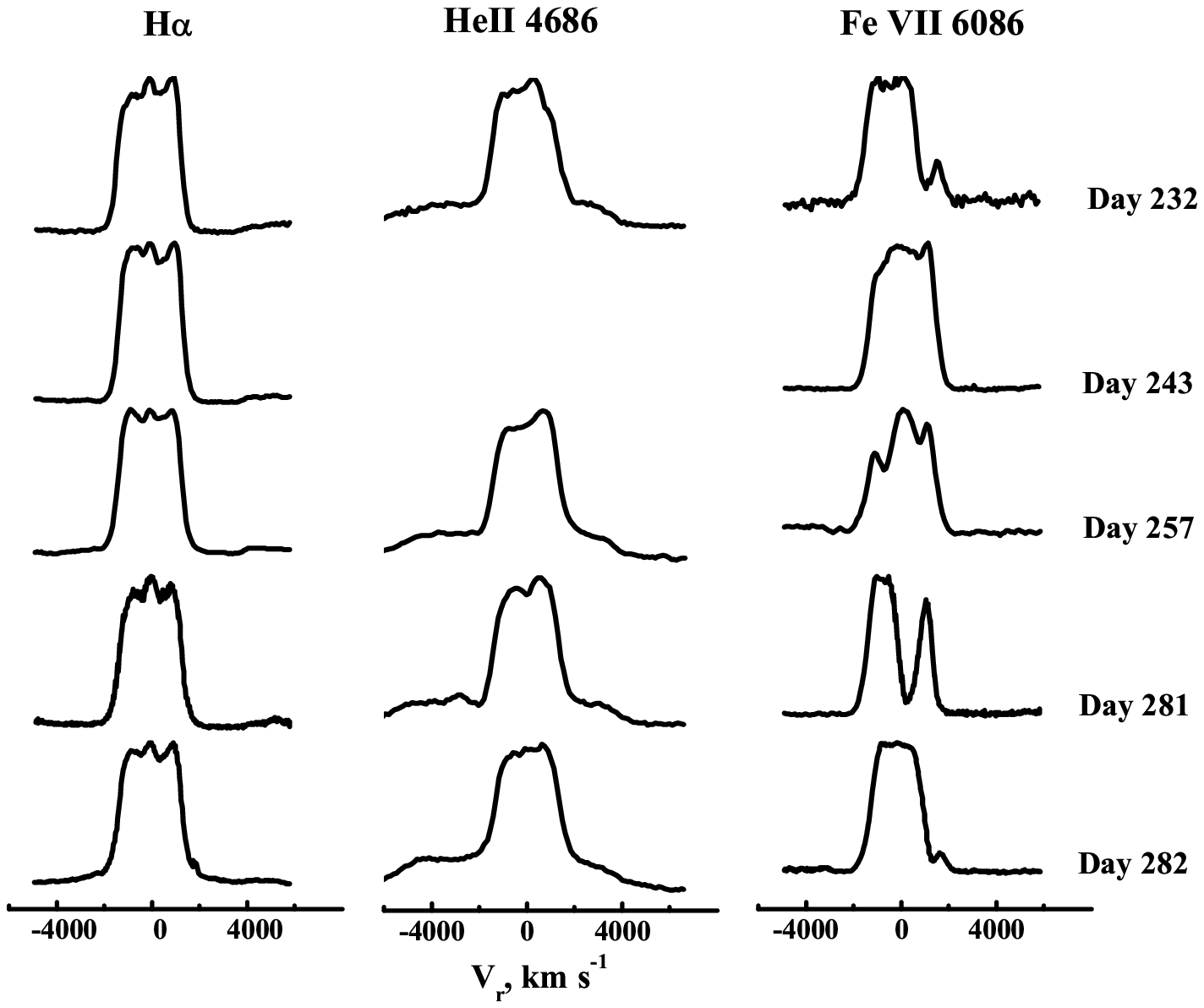}
\caption{Profiles of the H$\alpha$, HeII, and [FeVII] 6086 lines (days 232--282 after the brightness maximum). The line fluxes have been normalized to the maximum flux.}
\end{figure}

\begin{figure}
\hspace{-20mm}
\includegraphics[scale=0.8]{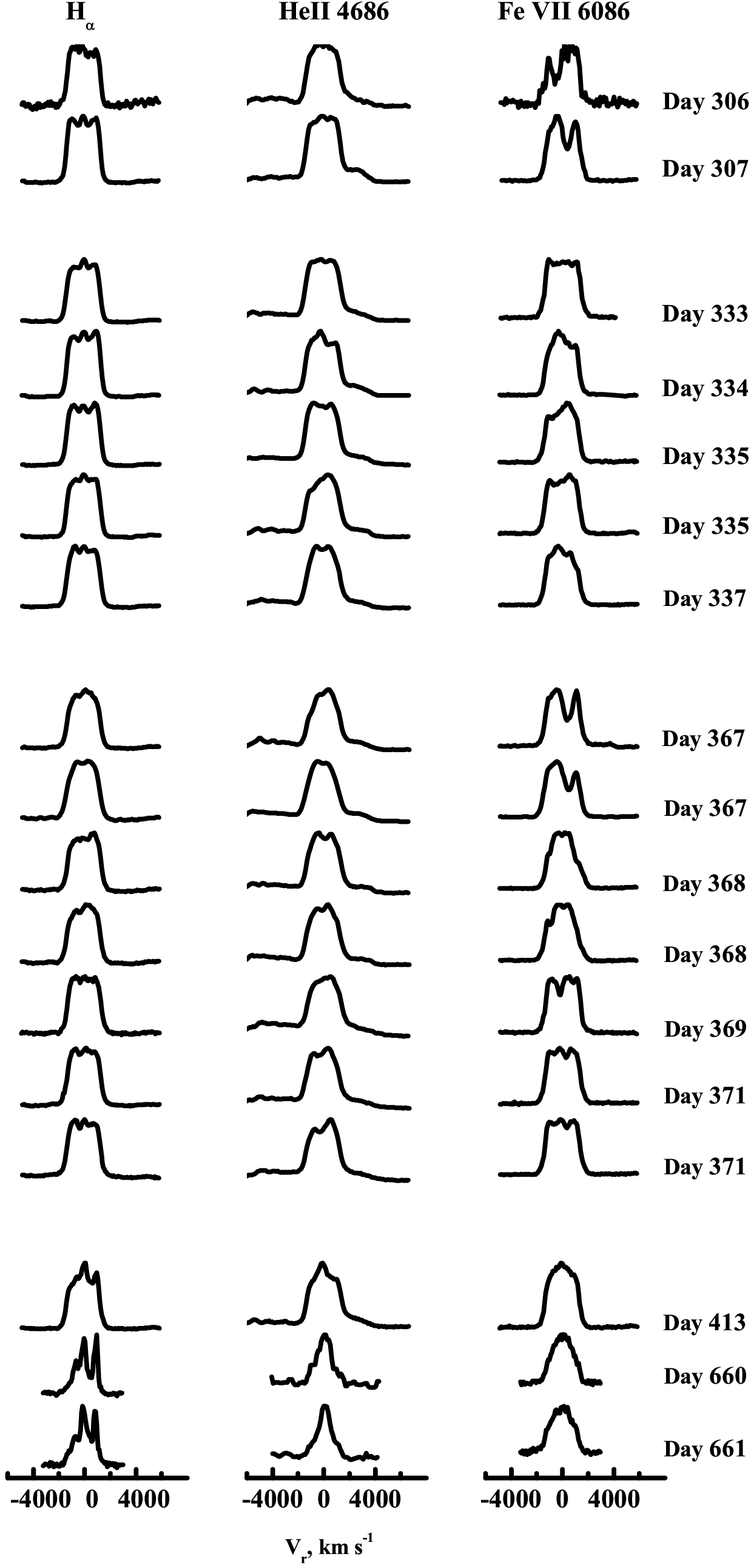}
\caption{Profiles of the H$\alpha$, HeII, and [FeVII] 6086 lines (days 306--661 after the brightness maximum). The line fluxes have been normalized to the maximum flux.}
\end{figure}

It may be that the fast (within a day) variations of the [FeVII] 6087 line are related to the orbital period; they can be interpreted in the same way as variations of the forbidden-line profiles observed in high-mass binaries consisting of a Wolf-Rayet star and an O star. Ignace et al. \cite{Ignace2009} suggested that the observed forbidden-line profile variations for such binaries could be explained as the effects of collisions of two winds: a high-velocity wind from the Wolf--Rayet star and a low-velocity wind from the O star. They modeled the profiles of several forbidden lines supposing that these lines were formed close to the collisional boundary of the two matter flows, and studied the dependence of the line-profile variations on the phase of the orbital period. They considered the following model for the interaction of the gas flows. Due to the wind from the Wolf-Rayet star flowing around the secondary, a conical cavity containing more rarified gas forms behind the O star. The orbital motion bends the conical cavity, creating a spiral. If spectral lines are formed close to the collisional boundary between the winds, we observe variations of the line profiles related to the orbital period. The lines formed in the regions adjacent to this cone exhibit the strongest variations.

It may be that the variations of the forbidden-line profiles of V458 Vul also result from collisions of matter flows from the nova with the matter of the planetary nebula within which the nova outburst occurred, expanding at a rate of about 20 km/s \cite{Wesson2008}. The gas flowing from the white dwarf travels around the secondary, forming a cavity of more rarified matter behind it. If the system's secondary is a red dwarf, and not a giant as in symbiotic novae, then noticeable profile-line variations should be observable within hours. Indeed, the orbital period was found by Rodriguez-Gil et al. \cite{Rodriguez-Gil2010} to be very short, 98 minutes.

The formation of a shock in the system due to the collision of two winds is supported by the detection of 0.6-10 keV X-rays \cite{Ness2009}. The hypothesis that the observed variations are due to collisions of two gas flows is also supported by the fact that the model profiles of forbidden lines in high-mass binary systems consisting of a Wolf-Rayet star and an O giant \cite{Ignace2009} bear a remarkable similarity to the [FeVII] line profiles of V458 Vul (days 333--337; Fig. 9).

We estimated the electron temperature and number density of the medium where the [FeVII] lines were formed from the ratio of the [FeVII] 3759 and 6086 lines, using the computations of Nussbaumer and Storey \cite{Nussbaumer1982}. We estimated the temperature to be about 120~000~K and the electron number density to be of order n$_{e}$ = 5$\times$10$^8$ cm$^{-3}$ on days 257--281 after the brightness maximum; by day 334, the temperature had decreased to 30 000 K and the electron number density to approximately n$_{e}$ = 5$\times$10$^7$ cm$^{-3}$.

Long-term (about 60-day) variations of the [Fe VII] 6087 line profile appeared during the 60-day brightening of V458 Vul (Fig. 7). We have no explanation for these variations of the [FeVII] 6087 line, but it seems likely that the brightness increase was related to the changing structure of the formation region of the [FeVII] lines.

\section{ABUNDANCES OF SELECTED ELEMENTS AND THE MASS OF THE V458 Vul ENVELOPE}

The abundances of chemical elements in novae can be estimated using methods for calculating the concentrations of several ions of a given element relative to the concentration of hydrogen. The total abundance is then the sum of the various ion abundances. The abundances of ions are determined from permitted recombination lines, in particular the HeII 4686 line, from the relation:

\begin{equation}
\frac{N(X^{i+})}{N(H^+)}=\frac{\lambda(X^{i+})}{\lambda(H\beta)}\frac{\alpha^{eff}(H\beta)}{\alpha^{eff}(\lambda)}\frac{I(\lambda)}{I(H\beta)},
\end{equation}

where $\alpha^{eff}(H\beta)$, $\alpha^{eff}(\lambda)$ are recombination coefficients, $I(\lambda)$, $I(H\beta)$ are the strength of the line used to derive the elemental abundance and the strength of the H$\beta$ line, and $H\beta$, $\lambda(X^{i+})$ and $\lambda(H\beta)$ are the wavelengths of the corresponding lines.

We estimated the helium abundance from the HeII 4686 line, which was one of the strongest lines in the nova's spectrum at the given phase. The relation used to calculate the concentration of doubly ionized helium was taken from Aller \cite{Aller1984}:

\begin{equation}
lg\frac{N(He^{++})}{N(H^+)}=-1.077+0.135\times lg(t_e)+\frac{0.135}{t_e}+lg(\frac{I(4686)}{I(H\beta)}),
\end{equation}

where $t_e=T_e/10^4 K$ and I (4686) is the observed intensity of the HeII 4686 line. 

The HeII 4686 line was blended with the [NeIV] 4721 line lying to the red. We applied a simple procedure to separate the lines in the blend. We divided the helium line into two halves at its center, then calculated the total flux as double the flux from the unblended left side of the line. The flux in the [NeIV] 4721 line was then taken to be the flux difference between the blend and twice the flux from the left half of the helium line. Since HeI lines were already absent on the date in question, we adopted the abundance of HeII for the lower limit of the total abundance of helium with respect to hydrogen, which our computations show to be N(He)/N(H)=0.44.

It is relatively easy to estimate the abundances of elements relative to hydrogen if the envelope is transparent to light from the white dwarf. Forbidden lines become very strong at this phase, called the nebular phase, and it is these lines that are used to determine the abundances of ions of the corresponding elements. The brightness fades smoothly during the nebular phase, and all the line profiles become essentially the same. However, V458~Vul exhibited a considerable long-term brightening by 1.5$^{m}$ followed by a brightness drop during its nebular phase. Moreover, the profiles of different lines were not all the same, and varied during the time covered by our observations, as was demonstrated in the previous section. Thus, despite the appearance of forbidden lines, the envelope remained partially opaque to the white dwarf's light for a considerable time. However, the line profiles had become similar by day 413. There was some difference related to the superposition of the narrow profiles belonging to the knots of the planetary nebula on the wide profiles of the envelope lines (Fig. 9). Moreover, the brightness again fell to practically its lowest value, and showed no appeciable variations. Based on these observations, we proposed that the envelope had become transparent by this date, so that we could now estimate the abundances of various elements. We did not consider later dates: unfortunately, the spectra we obtained on those dates do not cover the necessary spectral range, and we do not have a complete set of spectral lines. The line fluxes during the nebular phase after correction for reddening are presented in Tables 2. The color excess E(B-V)=0.65 was taken from Wesson et al. \cite{Wesson2008}.

\begin{table}[t]
\centering{\bf Table 2. Fluxes in the spectral lines on the 232nd-662nd days after the outburst}\\
\label{meansp} 
\begin{tabular}{l|c|c|c|c|c|c|c|c|c|c}
\hline\hline
\multicolumn{11}{c}{Day after brightness maximum}\\ 
\hline
Line (\AA)                                 & 232 & 243 & 257 & 281-282 & 306-307 & 333-337 & 367-371 & 413 & 603 & 659-662 \\
\hline
\lbrack NeV\rbrack~3425.8                   &     &     & 29  &  37     &         &   16.6  &  19.2   & 15.0&     &         \\
\lbrack FeVII\rbrack~3586                   &     &     & 5.5 &  5.4    &         &         &  1.7    & 2.2 &     &         \\
\lbrack FeVII\rbrack~3759                   &     &     & 4.1 &  4.6    &         &         &  2.5    & 2.1 &     &         \\
OVI 3811+3834                               & 1.2 &     & 2.1 &  1.9    &         &  0.6    &  1.7    & 0.4 &     &         \\
\lbrack NeIII\rbrack~3868.7+H$_{8}$         & 10  &     & 4.5 &  11.2   &  2.7    &  1.5    &  1.3    & 1.2 &     &         \\
\lbrack NeIII\rbrack~3968.7+H$_{\epsilon}$  & 2.1 &     & 1.6 &  1.8    &  1.1    &  0.7    &  0.6    & 0.4 &     &         \\
H$_{\delta}$+                               &     &     &     &        &         &         &         &      &     &         \\
NIII~4097.3,4103.9                          &  2.3&     & 1.7 &  2.1    &  1.4    &  0.9    &  0.9    & 0.4 &     &         \\
\lbrack OIII\rbrack~4363.2+H$_{\gamma}$     & 4.3 &     & 2.4 &  2.8    &  2.1    &  1.8    &  1.4    & 1.1 &     &         \\
NII+NIII 4640                               & 1.4 &     & 1.6 &  2.1    &  1.1    &  0.8    &  1.4    & 0.3 &     &         \\
\lbrack NeIV\rbrack 4720.0                  & 0.9 &     & 1.0 &  1.1    &  0.7    &  0.5    &  0.5    & 0.15&     &         \\
HeII 4685.7                                 & 9.8 &     & 9.7 & 11.1    &  8.5    &   5.8   &   6.6   & 3.3 &     &         \\ 
H$_{\beta}$                                 & 3.6 &     & 2.4 & 2.6     &  2.1    &   1.5   &   1.5   & 0.8 &0.14 &  0.09   \\
\lbrack FeVII\rbrack~4942+                  &     &     &     &        &         &         &         &      &     &         \\
\lbrack FeVI\rbrack~4974                    & 4.6 &     & 3.4 & 3.9     &  3.4    &  2.5    &  2.3    & 2.0 & 0.3 & 0.3     \\
\lbrack FeVI\rbrack~5176.0                  &  1.6& 2.2 &  1.8& 1.7     & 2.2     &  1.2    &  1.4    &  0.7& 0.09&  0.06   \\
\lbrack CaV\rbrack~5308.9                   &  1.0& 1.64& 1.0 & 1.0     & 1.5     &  1.1    &  1.0    &  0.8& 0.15&  0.12  \\
HeII 5411.5                                 & 1.0 & 0.9 & 1.1 & 0.9     & 0.7     &  0.6    &  0.7    &  0.3&0.05 & 0.05    \\
\lbrack FeVII\rbrack~5721                   & 1.2 & 3.0 &     &  1.4    & 1.9     &  1.5    &  1.6    & 1.3 & 0.18&  0.11   \\
HeI  5875.6                                 & 1.5 & 1.6 &     &  0.5    & +/0.32  &  0.16   & 0.08    &     &     &         \\
\lbrack FeVII\rbrack~6085.5                 & 1.7 & 4.0 & 2.0 & 2.1     & 3.1     &  2.4    &  2.4    & 1.9 & 0.28&  0.18   \\
H$_{\alpha}$                                & 8   &  11 &  5.4& 4.3     & 4.7     &  3.2    &  3.1    & 2.0 & 0.4 &  0.24   \\
HeI 6678.2                                  & 0.4 & 0.4 & 0.3 &  0.2    & +/0.13  &  0.13   &  0.12   &     &     &         \\
HeI 7065.7                                  & 1.0 & 1.1 & 0.35&  0.28   & +       &         &         &     &     &         \\
\hline
\end{tabular}
\end{table}

We determined the abundances or number densities of ions relative to hydrogen from forbidden lines using the relation

\begin{equation}
\frac{N(X^i)}{N(H^+)}=\frac{I(\lambda)}{I(H(\beta))}\frac{j(H(\beta)}{j(\lambda)}, 
\end{equation}

where $I(\lambda)$, $I(H\beta))$ are the line intensities in the observed spectrum, $j(H(\beta)), j(\lambda)$ are the volume emission coeffcients for the $H(\beta)$ line and a forbidden line with wavelength $\lambda$.

In this case, we calculated the volume emission coefficient for the forbidden line with a given temperature and density, and then calculated the abundances of ions relative to hydrogen using the observed intensities of the forbidden lines, normalized to the intensity of $H(\beta)$. Note that determining elemental abundances encounters diffculties because there are no [NIII], [OIII], [OII], or [SII] lines that can be used to determine the electron temperature and density in the spectrum of V458 Vul. In some such cases, the mean electron temperature of the nova envelope in the nebular phase is assumed to be 11 000 K \cite{Snijders1990} and the electron density, for example, to be the critical density for the [OIII] 4363 line \cite{Arkhipova2000}, which is about 10$^7$~cm$^{-3}$. However, the absence of [OIII] lines in the spectrum most likely indicates that the envelope's electron temperature is higher than is usually assumed for novae in the nebular phase. 

Also, due to the limited set of spectral lines corresponding to the same element but in different ionization states, we can derive only lower limits for the total abundances of neon, iron, and helium, and we can obtain only the abundances of the Ca$^{4+}$ and Ar$^{4+}$ ions in the case of calcium and argon. We estimated the electron temperature and density using the relation between these parameters and the intensity ratio of the [FeVII] 3759 and 6087 lines derived by Nussbaumer and Storey \cite{Nussbaumer1982}. According to \cite{Nussbaumer1982}, a ratio close to unity, as we found, corresponds to an electron temperature of about 30 000 K and a number density of about 5$\times$10$^7$ cm$^{-3}$.

We calculated the abundances of neon and argon using the NEBULAR.IONIC code \cite{Shaw1995}, which calculates the volume emission coeffcient for a forbidden line for a given temperature and density, then calculates the abundances of ions relative to hydrogen using the observed intensities of forbidden lines normalized to the $H(\beta)$ intensity. We determined the abundance of Ne$^{3+}$ from the [NeIV] 4721 line and the abundance of Ne$^{4+}$ from the [NeV] 3426 line. The absence of [OIII] lines from the spectrum indicates that lines of doubly ionized neon, [NeIII], are also very weak or completely absent from the spectrum, since the ionization potentials of these ions are similar. We thus obtained a lower limit for the neon abundance as the sum of abundances of the Ne$^{3+}$ and Ne$^{4+}$ ions. The abundances of these ions and our estimate of the total abundance, 4.76$\times$10$^{-4}$ are given in Table 3. The abundance of Ar$^{4+}$ was calculated from the [ArV] 7006 line: 0.13$\times$10$^{-5}$, also given in Table 3. We adopted the relation for the abundance of Ca$^{4+}$, derived from the [CaV] 5309 line, from \cite{Aller1984}. The abundance of this ion with relative to hydrogen is 5.15$\times$10$^{-6}$.

We estimated the iron abundance using two techniques. The first was to compare the iron abundance relative to neon assuming that the ratio of the Fe$^{6+}$ abundance to the Ne$^{4+}$ abundance was close to the ratio of their total abundances. We used the expression relating the ratio of the Fe$^{6+}$ and Ne$^{4+}$ abundances to the intensity ratio of the [FeVII] 6087 and [NeV] 3426 lines obtained by Shields \cite{Shields1978}:
\begin{equation}
\frac{I([FeVII]6087)}{I([NeV]3426)}=0.35e^{1.71/t}\frac{N(Fe^{6+})}{N(Ne^{4+})},
\end{equation}
where I([FeVII]6087)/I([NeV]3426) is the line intensity ratio and N(Fe$^{6+}$)/N(Ne$^{4+}$) the ratio of the ion abundances. Thus, with the intensity ratio of 0.126 for the  [FeVII]~6087 and [NeV]~3426 lines on day 413 (Table 3) and taking the lower limit of the neon abundance relative to hydrogen to be the sum of the abundances of the Ne$^{3+}$ and Ne$^{4+}$ ions, 4.76$\times$10$^{-4}$, we find that the abundance of iron relative to hydrogen is N(Fe)/N(H)=0.203~N(Ne)/N(H)=9.67$\times$10$^{-5}$.

In the second method, we estimated the total iron abundance as the sum of abundances relative to iron of the Fe$^{5+}$ ion (derived from the [FeVI] 5176 line) and the Fe$^{6+}$ ion (derived from the [FeVII] 6087 line). For this purpose, we used the computations of the volume emission coeffcients for the [FeVI] 5176 and [FeVII] 6087 lines of Nussbaumer and Storey \cite{Nussbaumer1982, Nussbaumer1978}. The the volume coefficient for the [FeVI] 5176 line was taken for 20 000 K. The total iron abundance obtained using the second method, 11.36$\times$10$^{-5}$, is close to that from the first method. Table 3 presents the iron abundances derived using both methods; we adopted the mean value, 10.52$\times$10$^{-5}$. The total iron abundances derived using the two methods are separated by a slash in the Table. All the elemental abundances listed in Table~3 were derived relative to hydrogen. Table 3 presents the abundances of helium, neon, argon, calcium, and iron relative to the solar abundances.

\begin{table}[t]
\vspace{6mm}
\centering{\bf Table 3. The total abundances of neon and iron and the abundances of helium, calcium, and argon ions in the envelope of V458 Vul}
\vspace{5mm}
\begin{tabular}{|c|c|c|c|c|c|c|c|c|}
\hline\hline
He$^+$/H     &  Ne$^{3+}$/H & Ne$^{4+}$/H  & Ne/H         & Ar$^{4+}$/H  & Ca$^{4+}$   & Fe$^{5+}$/H  & Fe$^{6+}$/H    & Fe/H \\ 
             &  10$^{-4}$   & 10$^{-4}$    & 10$^{-4}$    & 10$^{-5}$    &  10$^{-6}$  & 10$^{-5}$  &10$^{-5}$         &10$^{-5}$ \\ 
 4648        &   4720       & 3426         &              & 7006         &     5309    &  5176      &  6087            &         \\
\hline
  0.44       &   1.94       & 2.82         & 4.76         & 0.13         &   5.15      &  4.25     & 7.11             & 9.67/11.36 \\
\hline
\end{tabular}
\end{table}

Elemental abundances were calculated earlier for V458 Vul by Tsujimoto et al. \cite{Tsujimoto2009}, but from X-ray spectra. The abundances we have derived differ from those found in \cite{Tsujimoto2009}. Tsujimoto et al. \cite{Tsujimoto2009} detected a deficiency of oxygen and iron, compared to other metals (Ne, Mg, Si, S), relative to the solar values. However, due to the absence of helium lines in the spectrum, their calculations assumed the helium abundance to be solar. As described above, our results show that the helium abundance is a factor of four higher than the solar abundance, and that iron is not deficient: its abundance is almost a factor of four higher than the solar value. It is possible that we also overestimated the flux in the [FeVII] 6087 line by not taking into account the fact that this line can be blended with the [CaV] 6087 line, with the latter's contribution to the line intensity possibly being considerable.

Hybrid novae are unfortunately a rare class of novae, and their elemental abundances have not been determined earlier. For this reason, we are not able to compare the elemental abundances of V458 Vul with those for other novae of the same type. V458 Vul stands out among all known novae in its very high helium abundance, which is a factor of 4.4 higher than the solar value. However, this is not an exception. A similarly high helium abundance was detected for the HeN nova V1370 Aql \cite{Andrea1994}: it is almost a factor of six higher than the solar value. Some HeN novae also exhibit high iron abundances. For example, iron abundances derived for V382 Vel and V1370 Aql are higher than the solar values by factors of 8 \cite{Augusto2003} and 60 \cite{Andrea1994}, respectively. According to our calculations, V458 Vul stands out in its low neon abundance: compared to other novae with high-mass white dwarfs, the neon abundance is higher than the solar value by only a factor of five. However, this is likewise not exceptional: the neon abundance derived for V382 Vel is also not very high, exceeding the solar value by only a factor of ten \cite{Augusto2003}.

Another nova displaying very weak oxygen lines in the nebular phase is V2214 Oph \cite{Williams1991}.  Note that this nova's nebular-phase spectrum strongly resembles that of V458 Vul: the oxygen lines are absent or very weak, and, in addition to H$_{\alpha}$ and H$_{\beta}$, the HeII 4686 and [FeVII] 6087 lines are very strong.  However, V2214 Oph is an FeII nova \cite{Williams1991}. An enhanced helium abundance was derived for V2214 Oph, but it is not as high as for V458 Vul, the excess being only a factor of two \cite{Andrea1994}.  In addition, V2214 Oph exhibited a much higher iron abundance (exceeding the solar level by a factor of 24 \cite{Andrea1994}) than V458 Vul. Note that the FeII nova V2214 Oph also has a higher neon abundance than to V458 Vul, exceeding the solar value by a factor of 25 \cite{Andrea1994}.	

We determined the mass of the hydrogen envelope to be $M=n_{e}m_{H}V$ , where  $m_{H}=1.64\times10^{-24}$ g is the mass of the hydrogen atom and V the volume of the envelope. We obtained the volume of the envelope by comparing the nova's observed and theoretical luminosities: 4$\pi$$ D^{2}$F$_{H\beta}=4\pi \epsilon_{H\beta}V$, where D is the distance to the nova, $F_{H\beta}=0.08\times10^{-11}~erg~ cm^{-2}s^{-1}$ the $H_{\beta}$ flux, and $\epsilon_{H(\beta)}$  the $H_{\beta}$ volume emission coefficient.  We thus have									

\begin{equation}
M=7.9F_{H\beta}D^{2}t_{e}^{0.85}n^{-1}_{e}M_{\odot},
\end{equation}

As when estimating the elemental abundances, we used the $H_{\beta}$  flux measured on day 413 after the brightness maximum, the distance to the nova D=12.7 kpc, the temperature of the envelope normalized to  10$^4$, t$_{e}$=3,	and	the electron density	$n_{e}=5\times10^7~cm^{-3}$.  This yielded an envelope mass of $5.2\times10^{-6}M_{\odot}$. The envelope mass with helium taken into
account  is  $M(tot)=n_{e}m_{H}V(1+4*N(He)/N(H))$, where N(He)/N(H)=0.44 is the helium abundance relative to hydrogen. Then, $M(tot)=2.76n_{e}m_{H}V$ or $M(tot)=1.4\times10^{-5}M_{\odot}$. The envelope mass found by Rajabi et al. \cite{Rajabi2012}, $M\approx 4\times10^{-4}M_{\odot}$, is more than an order of magnitude higher than our value.
Table 4 presents all the main photometric, spectroscopic, and physical parameters of V458 Vul.
\begin{table}[t]
\vspace{6mm}
\centering{\bf Table 4. Main parameters of V458 Vul}\\
\label{meansp} 
\vspace{5mm}
\begin{tabular}{lc}
\hline\hline
$JD_{max}$                                         &$\approx$ 2454321.9\\
$V_{max}$                                          &8.14$^m$  \\
$M_{Vmax}$                                         &-8.7, -9.2 \\
Velocity class  (t$_{2}$=7, t$_{3}$=18)            &Fast\\
Distance, kpc                                      & 10.1, 12.7\\
Color excess, E(B-V)                               &0.65\\
Mass of the white dwarf, M$_{\odot}$             &$\approx$1.3 \\
``Spectral type'' 	                               &hybrid       \\
White dwarf's type                                 &ONe   \\
Envelope expansion velocity, km/s                  &1300-1560 km/c \\
Elemental abundances compared to the               &   \\
Sun (concentrations):                              &    \\
He/He$_{\odot}$                                    & 4.4   \\
Ne/Ne$_{\odot}$                                    & 4.8   \\
Fe/Fe$_{\odot}$                                    & 3.7     \\
Envelope mass, M$_{\odot}$                         & 1.4$\times$10$^{-5}$    \\
\hline
\end{tabular}
\end{table}

\section{PLANETARY NEBULA}

The  planetary  nebula  inside  which  V458  Vul erupted was discovered by Wesson et al. \cite{Wesson2008}, who found  bright  features  situated  symmetrically  with respect to the central star in the image of the planetary nebula, whch they ``knots''. Goranskij et al. \cite{Goranskij2010} also reported the detection of knots in the vicinity of the nova.	The slit of our spectrograph was positioned so that light from both the star and the knots was received.	Figure  10  displays  a  two-dimensional image of the spectrum of the nova and knots taken on June 30, 2009; Fig. 11 shows 1D spectra of the Northwestern  knot  (upper)  and  the  Southeastern knot (lower) in the planetary nebula. The spectra of the knots in Fig. 11 were obtained by summing all the spectra taken in 2009 (April 2, May 28-31, June 28 and 30, August 26, and September 24). The lines in the spectrum corresponding to the knots themselves are the [OIII] 4959, 5007, [NII] 5755, 6548, 6584, [SII] 6717, 6731 and [NeIII] 3869, 3968 lines. Other forbidden lines are mainly emission lines of the night sky. Moreover, the lower spectrum, corresponding to the Southeastern knot, contains broad lines from the
nova's spectrum.

We  estimated  the  electron  temperature  of  the knots from the forbidden nitrogen lines, [NII], and the  electron  number  density  from  the  forbidden sulfur lines, [SII], using the NEBULAR.TEMDEN
code \cite{Shaw1995}. The observed emission-line fluxes for the Northwestern and Southeastern knots of the planetary nebula, averaged for several dates and normalized to the  $H_{\beta}$ flux, are presented in Table 5. The fluxes were normalized so that the flux in the  $H_{\beta}$ line was $F(H_{\beta})$=100 and were calculated taking into account the reddening, \mbox{E(B-V)=0.65}. We obtained a temperature of about T$_{e}$=10 000 K and an electron density of about n$_{e}$=600~cm$^{-3}$ for the Northwestern knot, and a temperature of about T$_{e}$=13 000 K and an electron density of about n$_{e}$=750 cm$^{-3}$ for the Southeastern knot. The electron densities we derived for the knots are more than a factor of four higher than the electron density for the planetary nebula (n$_{e}$=155 cm$^{-3}$) \cite{Wesson2008}.

\begin{table}[t]
\vspace{6mm}
\centering{\bf Table 5. The observed emission-line fluxes, normalized to the H$_{\beta}$ flux for the Northeastern and Southwestern knots of the planetary nebula.}\\
\label{meansp} 
\vspace{5mm}
\begin{tabular}{|l|c|c|}
\hline\hline
Line                    & Northwestern    & Southeastern \\
                         & knot            &   knot   \\
\hline
\lbrack OIII\lbrack 4959 & 125      &    140 \\
\lbrack OIII\lbrack 5007 & 417      &    429 \\
\lbrack NII\lbrack 5755  & 8        &    9 \\
\lbrack NII\lbrack 6548  & 86       &    74 \\
H$\alpha$                & 286      &    309  \\
\lbrack NII\lbrack 6584  & 250      &    231  \\
\lbrack SII\lbrack 6716  & 15       &    9  \\
\lbrack SII\lbrack 6731  & 17       &    11  \\
\hline
\end{tabular}
\end{table}

\begin{figure}
\hspace{-10mm}
\includegraphics[scale=0.6]{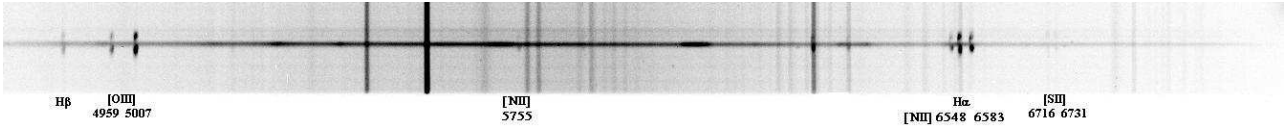}
\caption{Two-dimensional image of the spectrum of the Northwestern and Southeastern knots of the planetary nebula.}
\end{figure}

\begin{figure}
\hspace{-20mm}
\includegraphics[scale=0.9]{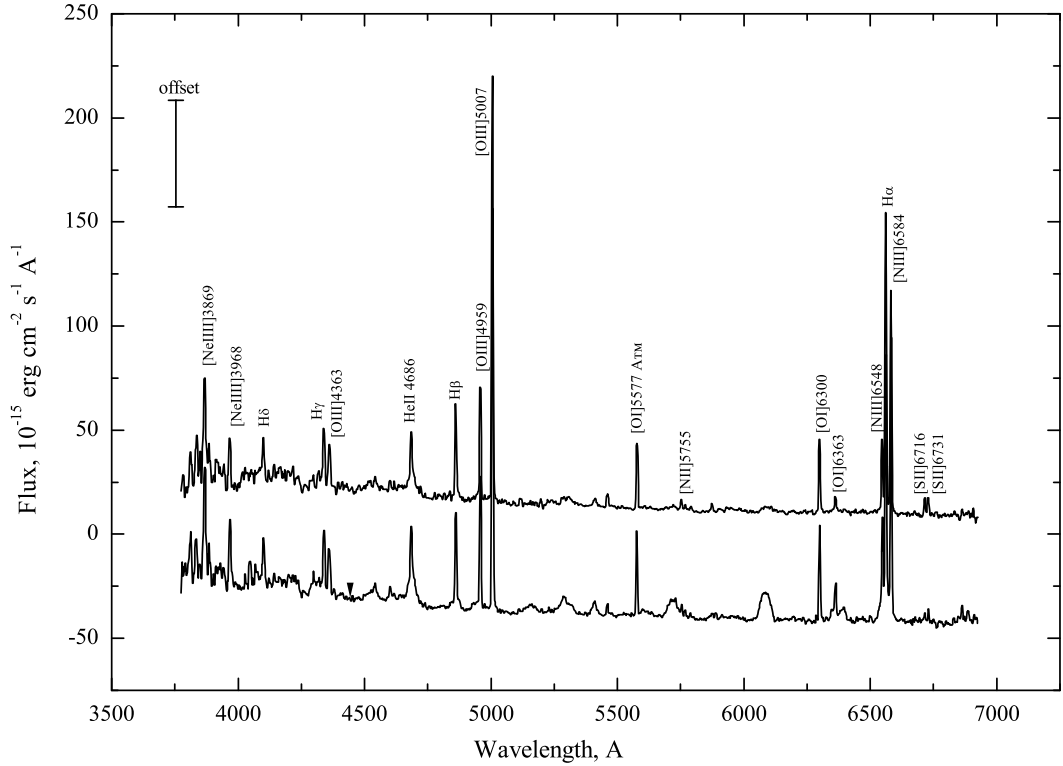}
\caption{The spectra of the Northwestern (upper curve) and Southeastern (lower curve) knots of the planetary nebula.}
\end{figure}

\section{Acknowledgments}
We are grateful to all variable-star observers who contributed to the AAVSO world database and whose observations were used in our study.

\end {document}